\documentclass[final,5p,twocolumn,times,number]{elsarticle}

\usepackage{amssymb}
\usepackage{lipsum}
\usepackage{lineno}
\usepackage{amsmath}

\usepackage{graphics}
\usepackage{graphicx}
\usepackage{epsfig}
\usepackage{dcolumn}
\usepackage{bm}
\usepackage{multirow}

\begin{document}

\begin{frontmatter}


\title{Indian participation in the construction of the Facility for Antiproton and Ion Research (FAIR) at Darmstadt, Germany} %% Article title

 \author[label1]{Saikat Biswas \thanks{Corresponding author}}
 \ead{saikat@jcbose.ac.in}

 \affiliation[label1]{organization={Department of Physical Sciences, Bose Institute},
             addressline={EN-80, Sector V},
             city={Kolkata},
             postcode={700091},
             state={West Bengal},
             country={India}}

%\author{} %% Author name

%\affiliation{organization={},%Department and Organization

%% Abstract
\begin{abstract}
India is a founder-member country to participate in the construction of the international multipurpose accelerator facility called the Facility for Antiproton and Ion Research (FAIR) at Darmstadt, Germany. Bose Institute, Kolkata, has been designated as the Indian shareholder of the FAIR GmbH and the nodal Indian Institution for co-ordinating Indian participation in the FAIR programme. 

Indian participation in FAIR is twofold. Firstly, the advancement of knowledge in nuclear astrophysics and reaction, high-energy nuclear physics, atomic \& plasma physics and application through the participation of Indian researchers, engineers and students in various experiments planned at FAIR. In addition to this, India is also contributing high-tech accelerator equipment as in-kind contribution to FAIR. 

Our active involvement include the designing, manufacturing and supply of in-kind accelerator items e.g. power converters, vacuum chamber, beam catchers, IT diagnostic cables among them and coordinating the participation of Indian scientists in the FAIR experiments including detector development, physics simulation, experimental data analysis.

Indian researchers have been participating in the two major experiments at FAIR, i.e. Nuclear Structure, Astrophysics and Reactions (NUSTAR) and Compressed Baryonic Matter (CBM) and in particular Bose Institute is involved in the CBM experiment, to study and characterize the matter created in the relativistic nucleus-nucleus collisions at high net baryon density and relatively moderate temperature.

In this article a brief overview on the FAIR facility, the experiments at FAIR and Indian participation are presented.

\end{abstract}

%%%%%%%%%%%%%%%%%%\begin{graphicalabstract}
%%%%%%%%%%%%%%%%%%\end{graphicalabstract}

%%%%%%%%%%%%%%%%%%\begin{highlights}
%%%%%%%%%%%%%%%%%%\item Research highlight 1
%%%%%%%%%%%%%%%%%%\item Research highlight 2
%%%%%%%%%%%%%%%%%%\end{highlights}

%% Keywords

\end{frontmatter}



%\linenumbers

\section{Introduction}
\label{sec1_intro}
The Facility for Antiproton and Ion Research (FAIR) is one of the largest international accelerator facilities, currently under construction at Darmstadt, Germany, for research on atomic and plasma physics (APPA), the high density nuclear matter (CBM), nuclear structure and astrophysics (NUSTAR), and basic physics on various areas around the weak and strong forces, exotic states of matter and the structure of hadrons (PANDA)  \cite{FAIR, FAIR_NIMA}. The location of FAIR is in the south of Frankfurt and just behind GSI Helmholtz centre, which is already well-known for the discovery of six elements of the periodic table namely Bohrium (Bh), Hassium (Hs), Meitnerium (Mt), Darmstadtium (Ds), Roentgenium (Rg), Copernicium (Cn).

The FAIR accelerator in its Modularised Start Version (MSV) is desired to be able to provide high intensity beams such as 1000 times the presently available primary heavy ion (HI) beam, 10000 times the radioactive ion beams and 100 times the antiproton beams. In this accelerator facility primary beams are expected to be 1-2 AGeV $^{238}$U$^{28+}$ at a rate of 5.10$^{11}$/s, 29 GeV Protons of rate 4·10$^{13}$/s, 10$^{10}$/s  $^{238}$U$^{92+}$ at 10 AGeV, $^{197}$Au$^{79+}$ at 11 AGeV, $^{40}$Ca$^{20+}$ at 14 AGeV along with secondary beams of rare isotopes of 1-2 AGeV and antiprotons up to 15 GeV  \cite{FAIR_Spiller_talk, QM2025_talk}. The prominent features of the FAIR accelerator are, highest beam intensities, brilliant beam quality, higher beam energies, highest beam power and four parallel operations.

\begin{figure}[htb!]
	\begin{center} 
 	\includegraphics[scale=0.41]{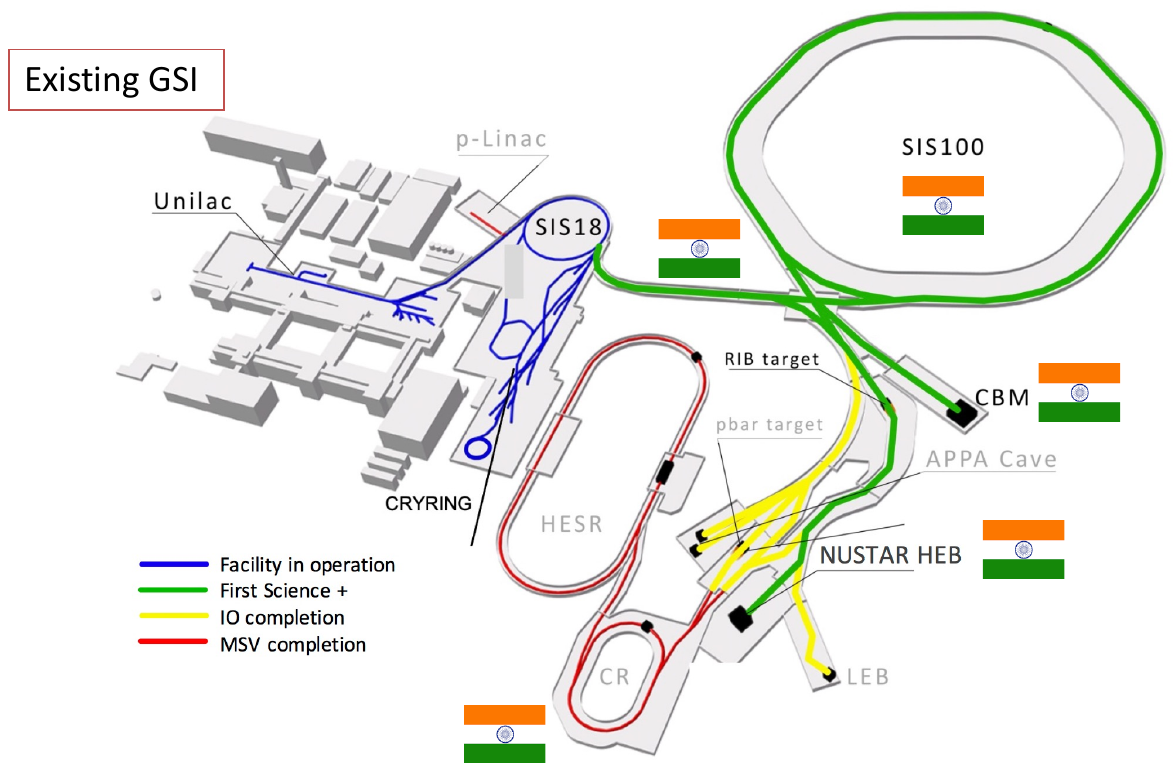}
 	\caption{(Colour online) The layout of the FAIR facility. }\label{fig_fair}
	\end{center}
 \end{figure}

The member states in FAIR are Germany, Finland, France, India, Poland, Romania, Russia, Sweden and Slovenia, whereas United Kingdom and Czech Republic are Associate Members. The layout of the FAIR accelerator facility along with the existing GSI is shown in Figure~\ref{fig_fair}, the mention of contributions from India are indicated with the Indian flag.
%%%%%%%%%%%%%%%%%%%%%%%%%%%%%%%%%%%%%%%%%%%%%%%%%%%
 %%%%%%%%%%%%%%%%%%%%%%%%%%%%%%%%%%%%%%%%%%%%%%%%%%%

Based on the availability of funds the implementation of an intermediate region of FAIR has been divided into various phases. The first one is the Early Science (ES) which is the FAIR pre-cursor programme at the Super-Fragment-Separator (S-FRS) and NUSTAR High-Energy Branch (HEB) served by beams from SIS18. Next one is the First Science (FS) program which is the first science at the S-FRS and NUSTAR High-Energy Branch (HEB) served by beams from the existing SIS100. Next comes the First Science + (FS+) which is in addition to FS the CBM branch served by beams from SIS100. Final step is the First Science ++ (FS++) where in addition to FS+ the branch into the APPA cave, and the NUSTAR Low-Energy Branch (LEB) will be there.

\begin{figure}[htb!]
	\begin{center} 
 	\includegraphics[scale=0.31]{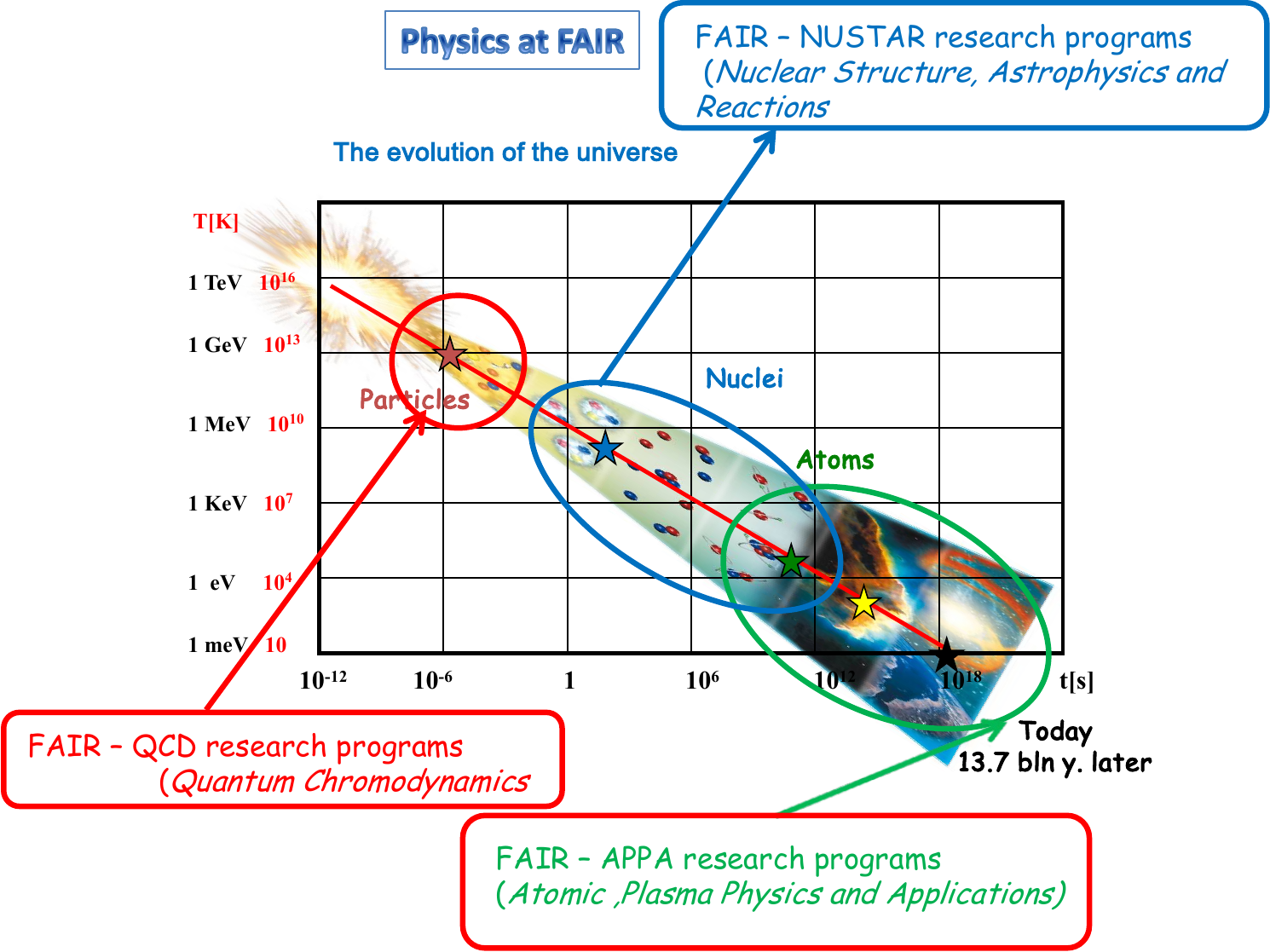}
 	\caption{(Colour online) Physics program at FAIR. }\label{fig_Physics}
	\end{center}
 \end{figure}

 The physics program at FAIR covers from very high energy to very low energy regions as happened after the Big Bang and depicted in Figure~\ref{fig_Physics}.
%%%%%%%%%%%%%%%%%%%%%%%%%%%%%%%%%%%%%%%%%%%%%%%%%%%
 %%%%%%%%%%%%%%%%%%%%%%%%%%%%%%%%%%%%%%%%%%%%%%%%%%%
The four experimental pillars in the physics program at FAIR have been shown i.e. APPA Physics (Atomic, Plasma Physics and Applications); CBM (Compressed Baryonic Matter); NUSTAR Physics (Nuclear Structure, Astrophysics and Reactions) and PANDA (Antiproton Annihilation at Darmstadt) \cite{FAIR_WFH}.

In the subsequent sections the the progress made towards FAIR project in synergy and the contribution from India will be discussed.

\section{Progress made towards FAIR project}
\label{progress}

The physical progress in FAIR project involves civil construction at the project site in Germany and design, development, fabrication of in-kind items (accelerator and detector) in various partner countries and Industries. In this report, we discuss about the equipment being made in India. The details of the progress made towards the project are described in the following subsections. 

\subsection{Civil Construction}
\label{civil}

%%%%%%%%%%%%%%%%%%%%%%%%%%%%%%%%%%%%%%%%%%%%%%%%%%%
 \begin{figure}[htb!]
	\begin{center} 
 	\includegraphics[scale=0.52]{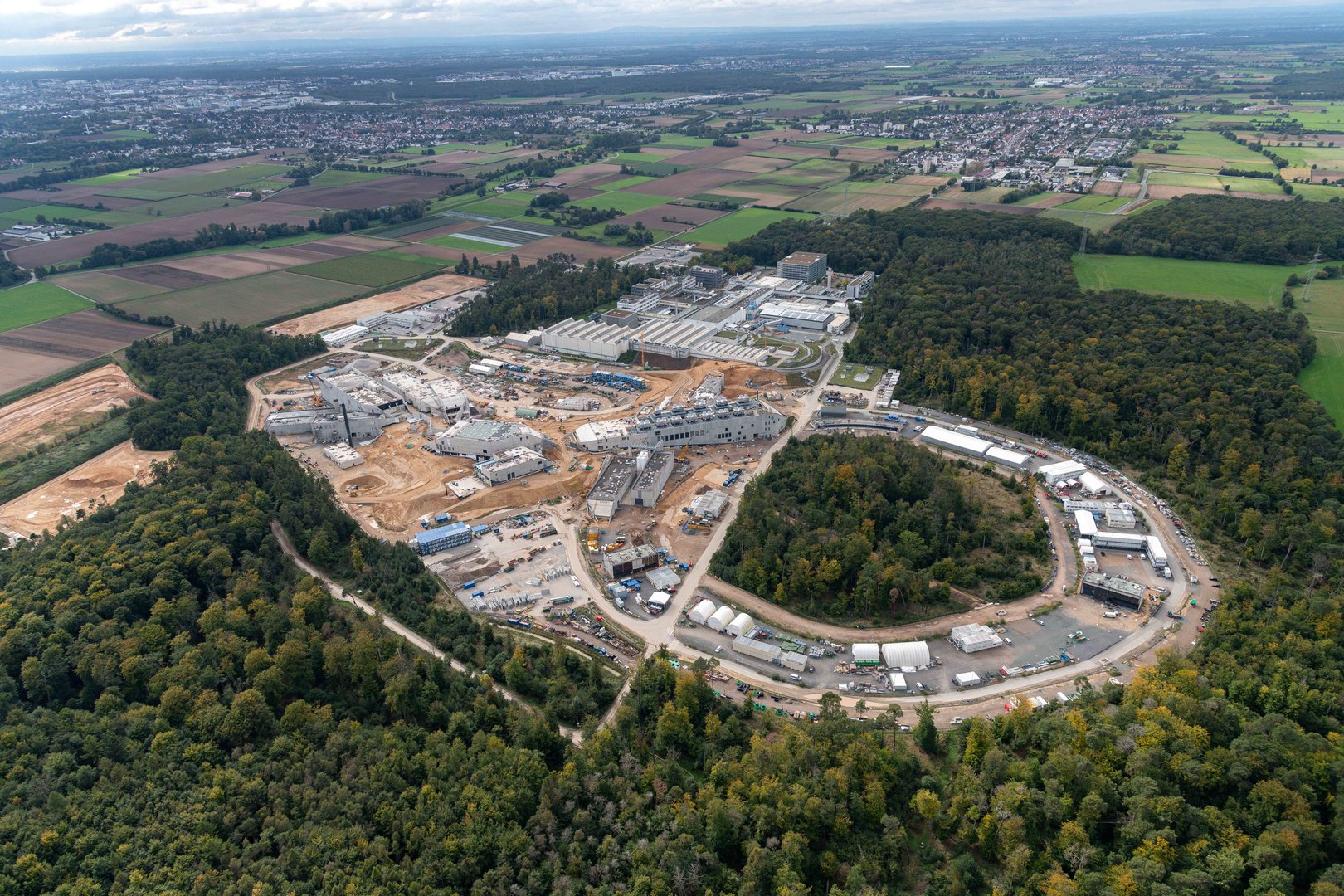}
 	\caption{(Colour online)  Civil construction of FAIR until September 2024 \cite{FAIR_media}. }\label{fig_civil}
	\end{center}
 \end{figure}
 %%%%%%%%%%%%%%%%%%%%%%%%%%%%%%%%%%%%%%%%%%%%%%%%%%%

\begin{figure}[htb!]
	\begin{center} 
 	\includegraphics[scale=0.55]{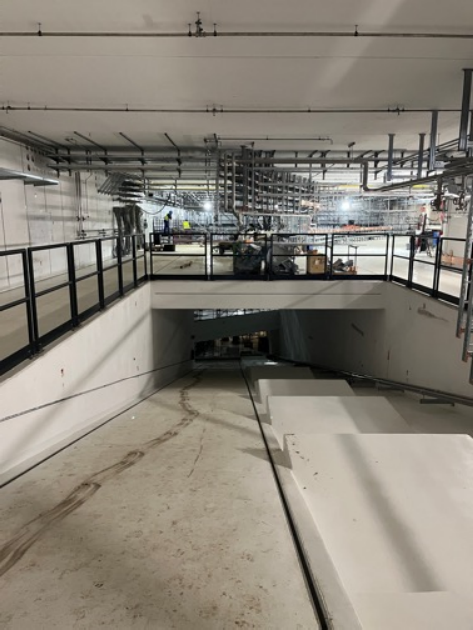}
 	\caption{(Colour online)  The way of the high energy beam transport (HEBT) line towards the SIS100 from SIS18.}\label{fig_sis100}
	\end{center}
 \end{figure}

\begin{figure}[htb!]
	\begin{center} 
 	\includegraphics[scale=0.65]{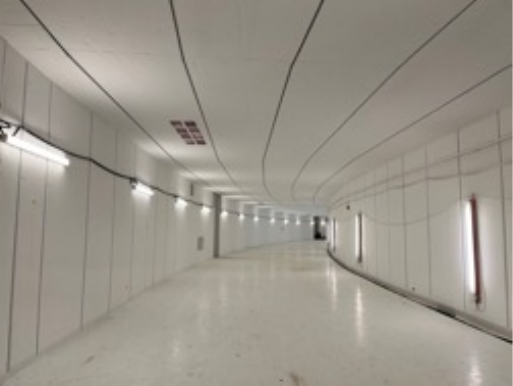}
 	\caption{(Colour online) Way of the 1100~m trip of beam under the roof. }\label{fig_1100}
	\end{center}
 \end{figure}

\begin{figure}[htb!]
	\begin{center} 
 	\includegraphics[scale=0.55]{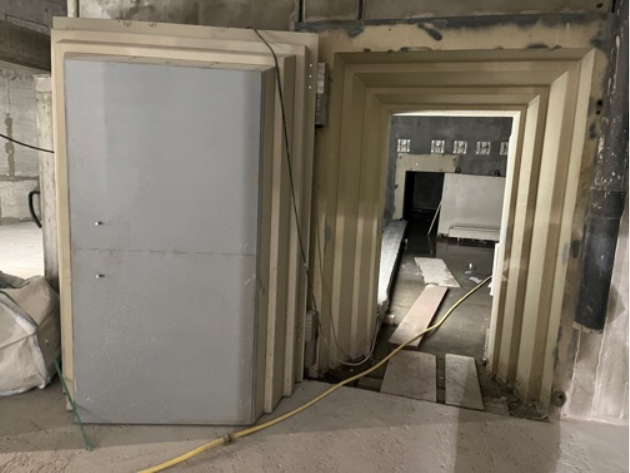}
 	\caption{(Colour online)  Very thick lead doors for radiation shielding. }\label{fig_lead}
	\end{center}
 \end{figure}

After initial delays in starting, the civil construction work initiated in 2017, is now almost complete. 
%The excavation work and the concrete work at the base of the tunnel of the entire main accelerator ring (SIS100) has been completed. The civil construction work for the first tunnel segment of SIS100 tunnel has also been completed. The civil construction work of one of an experimental cave (CBM), in which India has a big participation is going on at a fast speed. 
The main tunnel is complete in all respects along with the experimental cave of CBM and the high energy cave of NUSTAR. The accelerator equipment and devices like magnets, cryogens, power converters are currently  being installed.
The first beam from the accelerator for users in ES is expected by 2027. The status of the civil construction until September 2024 is shown in Figure~\ref{fig_civil}. The way of the proposed beam line towards the SIS100 is shown in Figure~\ref{fig_sis100} and the way of the 1100~m trip of beam in the tunnel is shown in Figure~\ref{fig_1100}. Among other components there are several very thick lead doors for radiation shielding, one of them in shown in Figure~\ref{fig_lead}.

%%%%%%%%%%%%%%%%%%%%%%%%%%%%%%%%%%%%%%%%%%%%%%%%%%%
 %%%%%%%%%%%%%%%%%%%%%%%%%%%%%%%%%%%%%%%%%%%%%%%%%%%
 
 %%%%%%%%%%%%%%%%%%%%%%%%%%%%%%%%%%%%%%%%%%%%%%%%%%%
 %%%%%%%%%%%%%%%%%%%%%%%%%%%%%%%%%%%%%%%%%%%%%%%%%%%
%%%%%%%%%%%%%%%%%%%%%%%%%%%%%%%%%%%%%%%%%%%%%%%%%%%
 %%%%%%%%%%%%%%%%%%%%%%%%%%%%%%%%%%%%%%%%%%%%%%%%%%%

\subsection{In-kind Contribution from India}
\label{inkind}

As per the present status, India's contribution include ultra-stable Power Converters, Ultra High Vacuum Chambers, Beam Catchers, IT diagnostic  Cables, Detectors related to experiments among others as in-kind contributions. India has made significant progress towards implementation of the following in-kind contributions.

\subsubsection{Low Energy Buncher magnets (LEB)}
\label{LEB}

Low Energy Buncher (LEB) of the  Superconducting Fragment Separator  (Super-FRS) is the device for  particle identification after secondary reactions. Dipole, quadrupole and sextupole  magnets forming the Energy Buncher  have to accept fragment beams  requiring large usable apertures with stringent magnetic field quality. 
%Initially, India has decided to supply superconducting magnets for LEB of FAIR as an in-kind item. However, after completion of its physics design and the basic engineering design by Variable Energy Cyclotron Centre (VECC), India, 
Engineers from Variable Energy Cyclotron Centre (VECC), Kolkata completed the physics and basic engineering design and subsequently completed the CAD modelling based on the engineering design. %its cost of production was found to be many-fold higher compared to the price estimated called the FAIR Cost-Book price. 
%As a result the production withdrawn from India although 
Based on the comparatively very high cost of production, the task of production in India is withdrawn. However, the Comprehensive Design Report (CDR) and Final Design Report (FDR) was cleared by FAIR before that. 
The design has been credited with a significant amount and as an offshoot, Indian engineers have been offered to be consultant for the dipoles.

\subsubsection{Ultra Stable Power Converters}
\label{USPC}

%%%%%%%%%%%%%%%%%%%%%%%%%%%%%%%%%%%%%%%%%%%%%%%%%%%
 \begin{figure}[htb!]
	\begin{center} 
 	\includegraphics[scale=0.85]{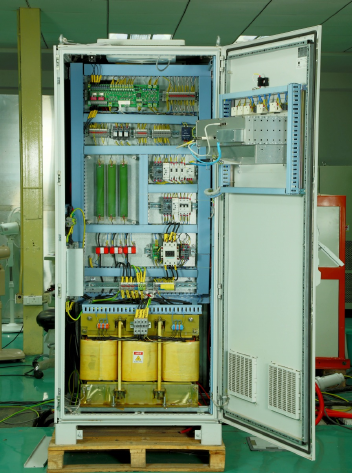}
 	\caption{(Colour online) Photograph of a power converter module supplied to FAIR. }\label{fig_PC}
	\end{center}
 \end{figure}
 %%%%%%%%%%%%%%%%%%%%%%%%%%%%%%%%%%%%%%%%%%%%%%%%%%%

The High Energy Beam Transport (HEBT) system of the FAIR accelerator facility provides a loss-free and emittance conserving transfer of ion, proton and antiproton beams to and from the synchrotrons and storage rings; to and from the Super-Fragment Separator; to the antiproton production target and separator; to the experimental areas (CBM and APPA) \cite{SR08}. The HEBT system of the FAIR consists of about 2.57~km of beam lines which require supply buildings and shielding. The beam line centre is placed typically 1.4~m above the floor and laterally decentred to allow transport of large items on one side of the beam pipe. Diversified beam parameters in HEBT lead to the requirement of both, room temperature as well as superconducting magnets. Power converters are required for powering these superconducting and room temperature magnets at the FAIR accelerator facility. Similarly power converters are required for the SIS100 magnets. The output from the power converter shall be supplied to the load (i.e. a quadrupole magnet or a steerer) through a coaxial cable. These power converters are ultra stable, i.e. ppm level stability in voltage and current is required. These power converters include single and dual power supplies in one rack.
So far in India total 454  power converter modules for SIS 100 and HEBT quadrupole and steering magnets are built by ECIL (Electronics Corporation of India Limited), Hyderabad and delivered to FAIR after detailed R\&D and tests. The photograph of the back side of a typical power converter module built at ECIL is shown in Figure~\ref{fig_PC}. A new set of 70 nos. of power converters for Corrector (Steerer, Sextupole and Octapole) Magnets in Super-FRS is under fabrication.

\subsubsection{Ultra-high Vacuum Chambers for beam diagnostics}
\label{UHVC}

%%%%%%%%%%%%%%%%%%%%%%%%%%%%%%%%%%%%%%%%%%%%%%%%%%%
 \begin{figure}[htb!]
	\begin{center} 
 	\includegraphics[scale=0.65]{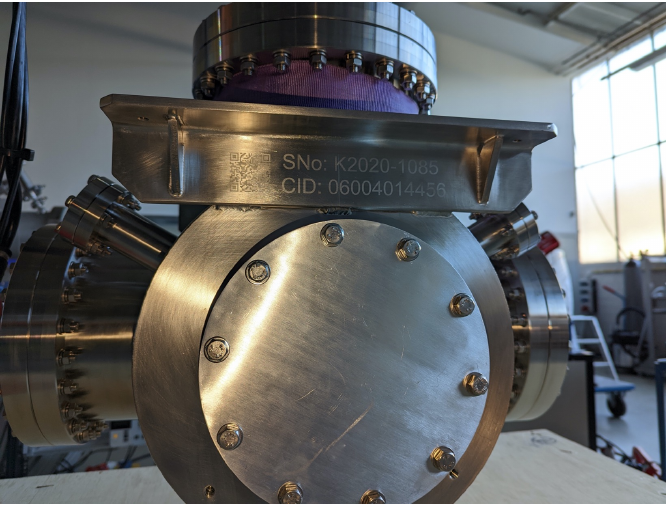}
 	\caption{(Colour online) Photograph of a vacuum chamber supplied to FAIR. }\label{fig_VC}
	\end{center}
 \end{figure}
 %%%%%%%%%%%%%%%%%%%%%%%%%%%%%%%%%%%%%%%%%%%%%%%%%%%

The diagnostic vacuum chambers in HEBT are the host for most of the beam diagnostic elements. The dimensions of diagnostic chamber depend on the aperture of the beam diagnostic detectors and the number of beam diagnostic elements in the chamber. The vacuum chambers are intended for ultrahigh vacuum (UHV) environments, up to 10$^{-8}$~mbar with working temperature of the chamber between 15-50$^\circ$C. The vacuum requirements of these chambers are defined by HEBT vacuum requirements e.g. the integral leak rate needs to be smaller than 1~$\times$~10$^{-10}$~mbar.l/s. Therefore, materials, manufacturing processes and cleaning procedures are needed to be as perfect as possible. The beam diagnostic Vacuum Chamber consist of top cover, chamber barrel, beam line entry and exit flange, planes for fiducial target seats, connections to alignment bridges and pipe sockets for diagnostic elements, vacuum generation and measurement. Each Diagnostic chamber needed to be mounted on a support frame and has to be aligned to beam axis. The vacuum chambers require use of highly cleaned and degassed weldable austenitic stainless steel with very low sulphur content (sulphur $<$ 0.015\% w/w) and low magnetic permeability (Relative Magnetic permeability $<$1.05). Welding and other manufacturing processes require special handling. Multistep cleaning is required to ensure vacuum quality etc. A major challenge is to weld up to 7 cylindrical ports on a barrel, which, when completed, should maintain mechanical tolerances at the level of tens of microns. The technology demands extremes care and quality control. The production of such vacuum chambers is done in collaboration with VTPL, IUAC, VECC and Bose Institute. 58 such chambers have been fabricated in a company in Bengaluru, India called the Vacuum Techniques Pvt. Ltd., Bengaluru, India and supplied to FAIR. All the chambers are accepted by FAIR after the site acceptance test. The photograph of a UHV chamber made in India is shown in Figure~\ref{fig_VC}.

\subsubsection{IT and Diagnostic Cables}
\label{IT}

%%%%%%%%%%%%%%%%%%%%%%%%%%%%%%%%%%%%%%%%%%%%%%%%%%%
 \begin{figure}[htb!]
	\begin{center} 
 	\includegraphics[scale=1.0]{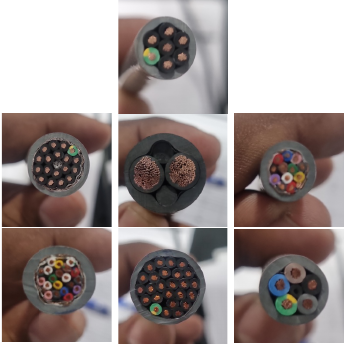}
 	\caption{(Colour online) Cross-sectional view of different IT cables.}\label{fig_IT}
	\end{center}
 \end{figure}
 %%%%%%%%%%%%%%%%%%%%%%%%%%%%%%%%%%%%%%%%%%%%%%%%%%%

The IT and Diagnostic Cables are required for signal or data transfer for diagnostic purpose. Fabrication of such cables needs radiation hard insulation as these cables will be used in radiation environment of FAIR accelerator facility. In these cables EBXL-XLPE insulation with halogen free sheath is used as per user requirement. 52 types of such cable of total length of about 930~km are fabricated by Siechem Technologies Pvt. Ltd., Chennai, India and delivered to FAIR in three lots. The cross-sectional view of some of the IT cables are shown in Figure~\ref{fig_IT}.  

\subsubsection{Beam Catchers (BC)}
\label{BC}
The Beam Catchers are required to safely catch the unreacted primary beams and a large share of unwanted fragments after the target in Super-FRS in the FAIR accelerator facility. These are primarily energy intercepting and dissipating devices. Beam Catchers are widely used for intercepting beams of various energy levels, starting from lower to very high energy beams from accelerators such as proton or antiproton and ion beams. The beam catchers are extremely important systems of large modern accelerator facilities both for basic and applied research as well as nuclear power applications. 

%%%%%%%%%%%%%%%%%%%%%%%%%%%%%%%%%%%%%%%%%%%%%%%%%%%
 \begin{figure}[htb!]
	\begin{center} 
 	\includegraphics[scale=0.65]{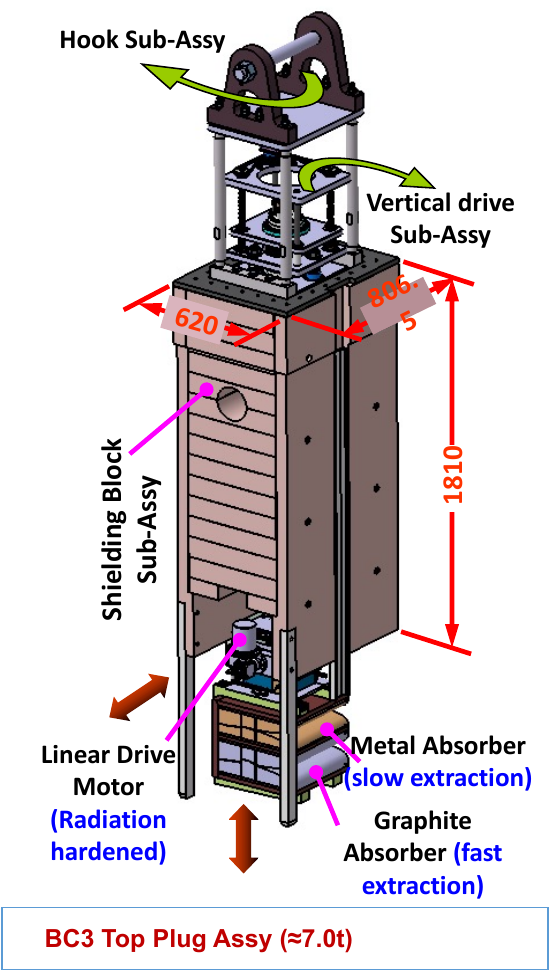}
 	\caption{(Colour online) Schematic of the plug of a beam catcher. The absorbers are at the bottom and movable both vertically and horizontally.}\label{fig_BC1}
	\end{center}
 \end{figure}
 %%%%%%%%%%%%%%%%%%%%%%%%%%%%%%%%%%%%%%%%%%%%%%%%%%%

Since beam catchers are to be designed to safely absorb and dissipate the kinetic energy of the particle, the primarily technological challenges are to distribute the huge thermal gradient compounded by thermal shockwaves and material irradiation damage. In addition to that, downstream ion-optical systems are also to be protected from radiation damage.

%%%%%%%%%%%%%%%%%%%%%%%%%%%%%%%%%%%%%%%%%%%%%%%%%%%
 \begin{figure}[htb!]
	\begin{center} 
 	\includegraphics[scale=0.5]{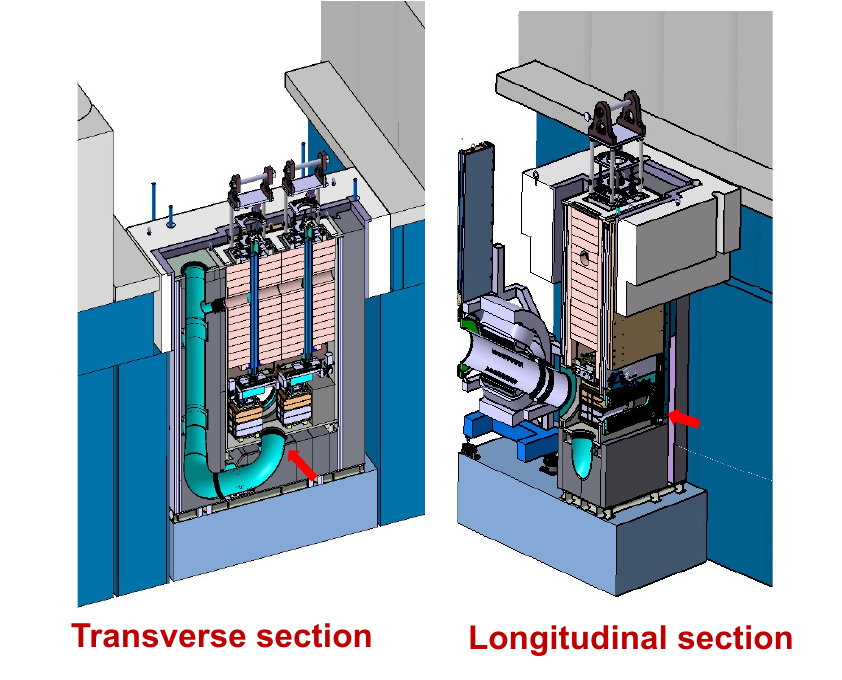}
 	\caption{(Colour online) Cross-sectional view of the transverse section (left) and longitudinal section (right) of a beam catcher. }\label{fig_BC2}
	\end{center}
 \end{figure}
 %%%%%%%%%%%%%%%%%%%%%%%%%%%%%%%%%%%%%%%%%%%%%%%%%%%

%%%%%%%%%%%%%%%%%%%%%%%%%%%%%%%%%%%%%%%%%%%%%%%%%%%
 \begin{figure}[htb!]
	\begin{center} 
 	\includegraphics[scale=0.032]{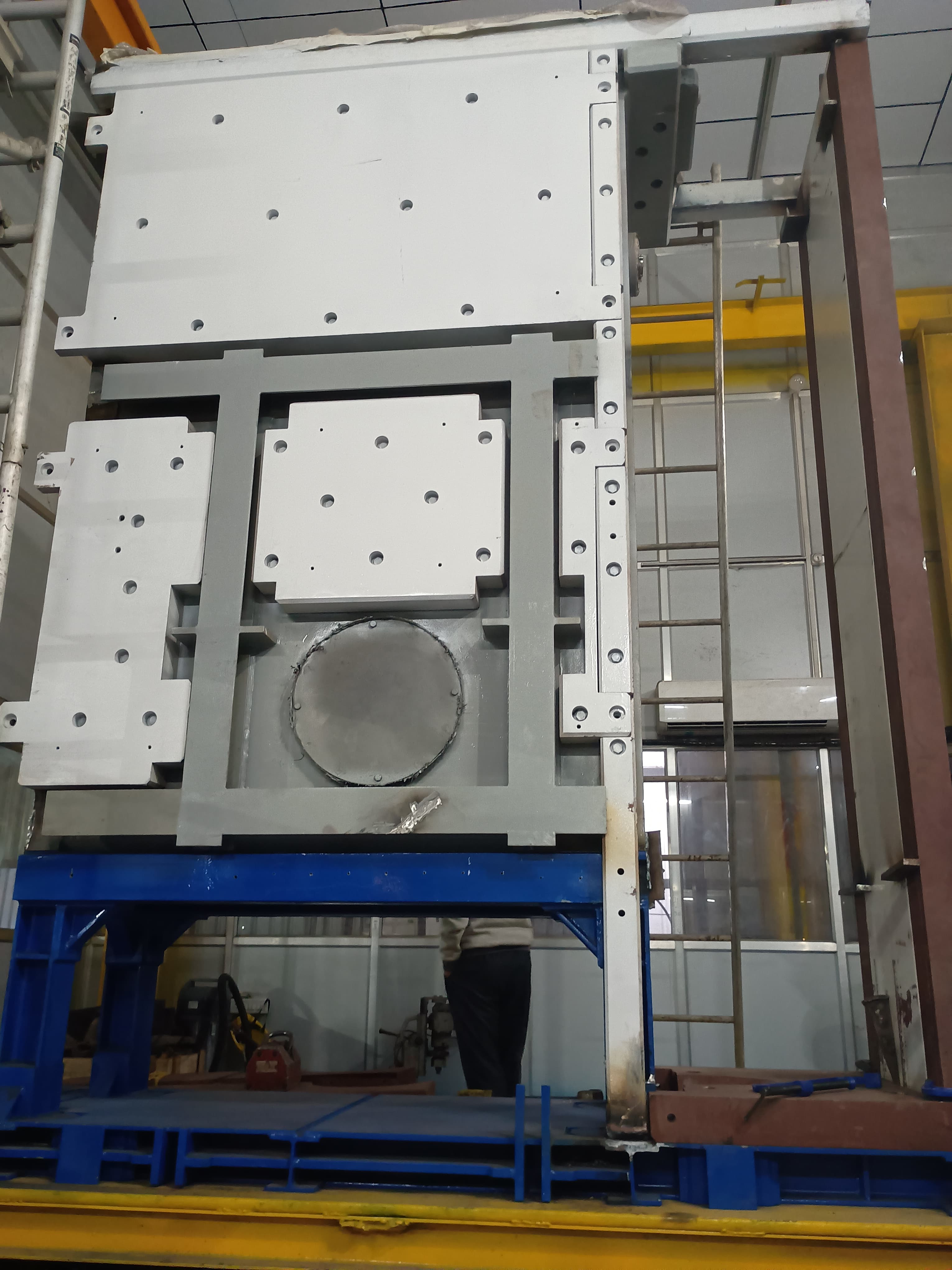}
	 	\includegraphics[scale=0.032]{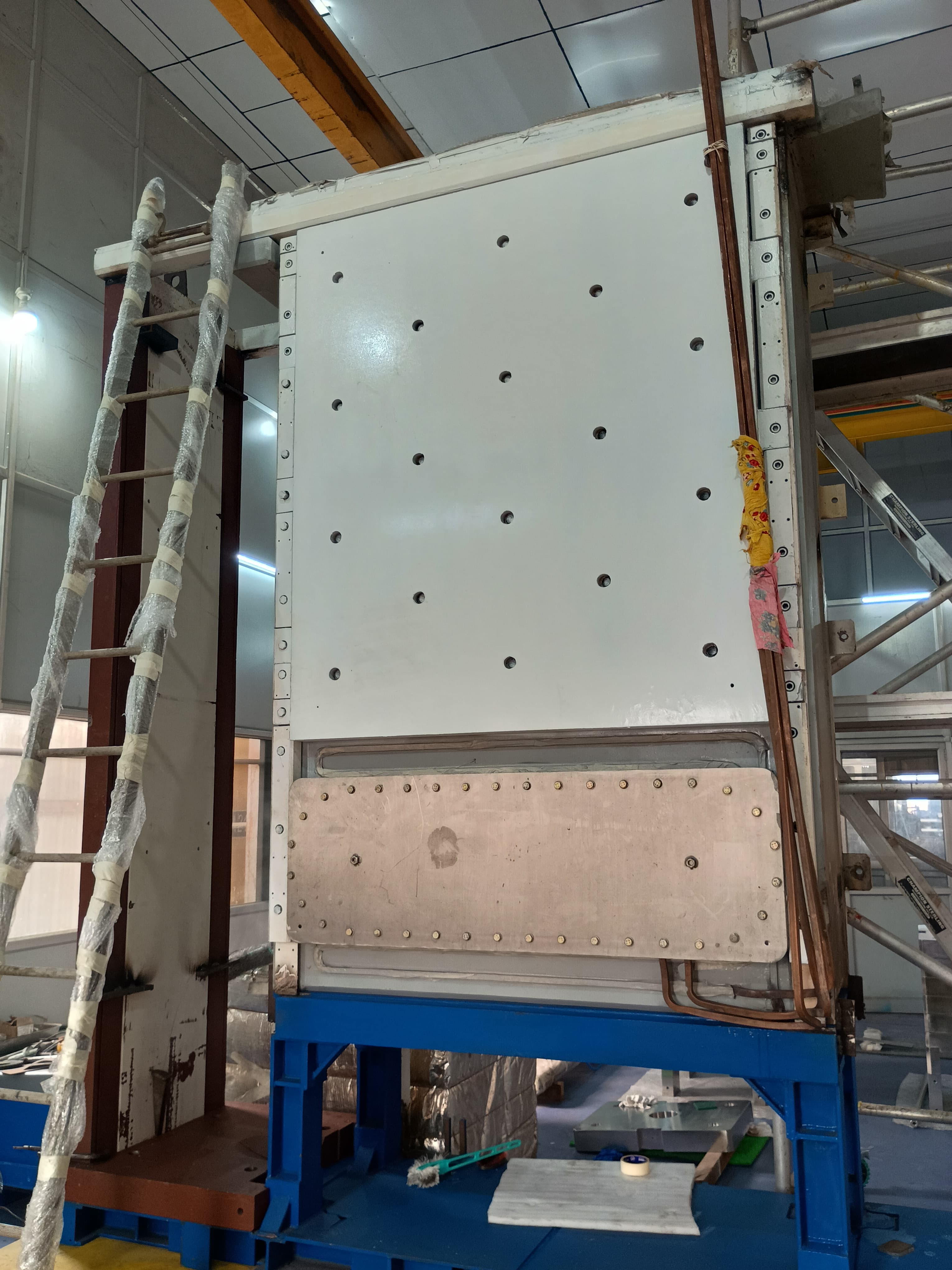}
 	\caption{(Colour online) Picture of the vacuum chamber of BC-3.}\label{fig_BSVC}
	\end{center}
 \end{figure}
 %%%%%%%%%%%%%%%%%%%%%%%%%%%%%%%%%%%%%%%%%%%%%%%%%%%

In Super-FRS three such beam catchers namely BC-1, BC-2 and BC-3 will be used. All three beam catchers will be placed in the Super-FRS tunnel, aligning with the beam-line and has the primary function of safely stopping unwanted fragments and the primary beam on requirement. All beam catcher chambers will contain two insertable shielding plugs to which a copper and a graphite beam absorbers are mounted for slow extraction and fast extraction mode of operation, respectively. A transmission drive for vertical positioning of the required absorber is used and the third beam catcher (BC-3) has the  additional provision of horizontal positioning of the absorbers for full or partial obstruction of beam. The absorbers are water cooled. The challenges to build this beam catchers are, (a) they must be integrated directly into the separator, (b) huge average power ($\sim$23~kW) dumps in very short time ($\sim$100~ns), (c) both fast and slow extraction method needs to be incorporated and (d) absorber is needed to be built suitable for remote handling. The dimension of the structural Frame (L$\times$B$\times$H) of each beam catcher is 2838~mm $\times$ 860~mm $\times$ 3350~mm whereas the the dimension of the vacuum cavity (L$\times$B$\times$H) is 1600~mm $\times$ 600~mm $\times$ 2520~mm. The schematic of the BC-3 is shown in Figure~\ref{fig_BC1} whereas the sectional views in the transverse and longitudinal directions are shown in Figure~\ref{fig_BC2}. Both sides of the vacuum chamber of BC-3, fabricated in India, are shown in Figure~\ref{fig_BSVC}.

The design of these beam catchers are done mainly by CSIR-CMERI in collaboration with VECC, Bose Institute and FAIR. The fabrication of these beam catchers is ongoing at Trident Auto Components Pvt. Ltd., Kanpur.

\section{Indian participation in FAIR experiments}
\label{expt}
Indian Institutes and Universities are at present involved in two major experiments at FAIR - NUSTAR and CBM. 
%NUSTAR experiment will help us to understand the stars by studying the behavior of atomic nuclei. We are all made of stardust. This is because stars and stellar explosions create the chemical elements of which our bodies and all living things are composed. 
With the help of the exotic rare ion beams (RIB), NUSTAR will study the nuclear astrophysical phenomena to understand the synthesis of elements. Only the simplest element, hydrogen, was created exclusively during the Big Bang, which also generated much of the next-lighter element, helium, as well as significant traces of the lightest metal, lithium. NUSTAR will also study the synthesis of heavier elements. 
%To understand the stars, we need to understand atomic nuclei, what the scientists strive for with the experiments of the NUSTAR collaboration. They want to study the nuclear reactions that occur inside stars. This leads us to the world of exotic isotopes helping us to understand the synthesis of the heavier elements in nature. 

\begin{figure}[htb!]
	\begin{center} 
 	\includegraphics[scale=0.6]{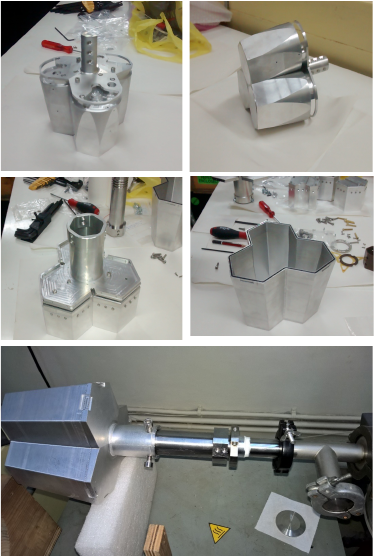}
 	\caption{(Colour online) Components of DEGAS detector of NUSTAR at FAIR.}\label{fig_nustar}
	\end{center}
 \end{figure}

Indian scientists are building the components of the Decay Spectroscopy (DESPEC) Germanium Array Spectrometer (DEGAS) detector set up to be employed at the Low Energy Branch (LEB) of the Super Fragment Separator (Super-FRS) at FAIR \cite{AK_nustar, AS_nustar}. Another group of Indian scientists are developing the Modular Neutron Spectrometer (MONSTER) to detect highly energetic neutrons in the reaction \cite{AR_nustar, KB_nustar, DV_nustar}. A third Indian group is working on the precision Measurements using Advanced Trapping System (MATS) aiming high precision mass measurements of short-lived radio nuclides. Some components of the DEGAS spectrometer are shown in Figure~\ref{fig_nustar}.

%%%%%%%%%%%%%%%%%%%%%%%%%%%%%%%%%%%%%%%%%%%%%%%%%%%
 %%%%%%%%%%%%%%%%%%%%%%%%%%%%%%%%%%%%%%%%%%%%%%%%%%%

The CBM experiment will create and study the conditions inside a supermassive object such as neutron star for a split second by colliding atomic nuclei at relativistic speeds. When a massive star reaches the end of its life, it explodes as a huge supernova, leaving behind an incredibly dense central core - a neutron star. 
%Although it is only the diameter of a city, it weighs around one million times more than the whole Earth. 
Researchers aim to use the CBM experiment to find out the behaviour of the strongly interacting matter under such extreme densities.

\begin{figure}[htb!]
	\begin{center} 
 	\includegraphics[scale=0.23]{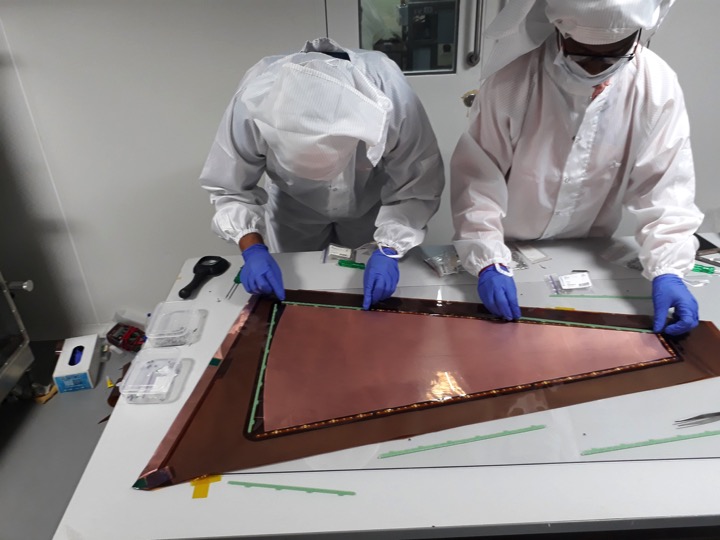}
 	\caption{(Colour online) Fabrication of a GEM chamber for CBM-Muon Chamber.}\label{fig_much}
	\end{center}
 \end{figure}

Indian researchers are building the Muon detection system for the CBM experiment. This detector system will be used to detect the short lived particles by detecting muon pairs as their decay
products along with the di-muon continuum. The detector system will have slices of graphite and iron plates of varying thickness as absorbers for the unwanted particles. Detector stations will be inserted in between each absorber pair. Different technologies will be used for different stations depending on the density of particles falling on them. The stations close to the interaction point will use Gas Electron Multiplier (GEM) technology due to their suitability of handling a high particle density \cite{gem_AKD, gem_SC, gem_SM}. The stations further will be using Resistive Plate Chamber technology \cite{rpc_sc, rpc_as} or Straw tube \cite{st_sr}. All detector stations will be part of India’s contribution to CBM. A picture of the fabrication of a CBM-Muon GEM Chamber is shown in Figure~\ref{fig_much}.

%%%%%%%%%%%%%%%%%%%%%%%%%%%%%%%%%%%%%%%%%%%%%%%%%%%
 %%%%%%%%%%%%%%%%%%%%%%%%%%%%%%%%%%%%%%%%%%%%%%%%%%%

\begin{figure}[htb!]
	\begin{center} 
 	\includegraphics[scale=0.55]{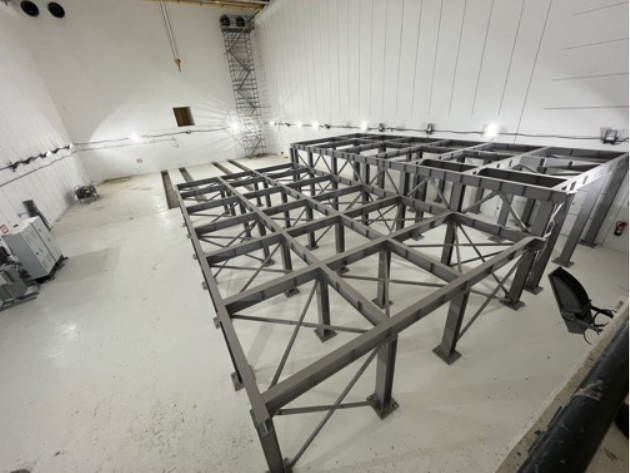}
 	\caption{(Colour online) Platform for the CBM experimental set-up at FAIR.}\label{fig_cbm}
	\end{center}
 \end{figure}

R\&D for detectors and electronics for India’s participation in two major experiments i.e., CBM and NUSTAR have been almost completed. One of the base structures to host the CBM experimental set-up is already in place at the CBM-cave as shown in Figure~\ref{fig_cbm}.

%%%%%%%%%%%%%%%%%%%%%%%%%%%%%%%%%%%%%%%%%%%%%%%%%%%
 %%%%%%%%%%%%%%%%%%%%%%%%%%%%%%%%%%%%%%%%%%%%%%%%%%%

%\vspace*{-0.8cm}
\section{Summary}
\label{sec_sum}
Mega projects like FAIR involve very advanced, unknown and complex technologies which are yet to be developed. The work is extremely challenging and is done by pooling all resources from all across the globe. Also, these projects involve inherent uncertainties which are beyond anybody’s control.

The civil construction work is almost completed at the project site in Germany. Technical equipment (in-kind items) building in India is also ongoing which include R\&D, prototyping, fabrication and shipment. The in-kind items are of two types; accelerator components; and advanced detectors and electronics for experiments. Both these types of equipment involve highly advanced technologies to be developed first and then, these in-kind items with very stringent technical specifications and quality control are to be manufactured by Indian industries. Some of the in-kind items from India are already delivered to FAIR and successfully passed the site acceptance tests, some other components are in advanced stage.

Indian researchers are also gearing up to install the detectors in the NUSTAR and CBM experiments. The R\&D for detectors and electronics for both the experiments are almost completed. The production of the real detectors are also started for some cases. India along with other countries are getting ready to take data when the first beam will come, expected in 2027.

%\vspace*{-0.6cm}
\section{Acknowledgements}
	
The author would like to thank Prof. Kaustuv Sanyal, Director, Bose Institute and Indian Shareholder and Dr. Praveen K Somasundaram, Head IC Division, Department of Science and Technology and Member of the FAIR Council for their support. The author would also like to thank the former Director of Bose Institute, Prof. Sibaji Raha, former Program Director of Indo-FAIR Coordination Centre (IFCC) Prof. Subhasis Chattopadhyay, former Associate Program Director, IFCC, Prof. Sanjay Kumar Ghosh, present Program Director, of Indo-FAIR project, Prof. Supriya Das and Dr. Inti Lehmann of FAIR for valuable suggestions. The author would like to thank Department of Science and Technology, Govt. of India and Department of Atomic Energy, Govt. of India for the funding of the project titled "India’s participation in the construction of the Facility for Antiproton and Ion Research (FAIR) at Darmstadt, Germany" (SR/MF/PS-02/2020 (E-6133) dt 08.10.2021).

\end{document}